# Teaching an Old Ball New Tricks
## Another Look at Energetics, Motion Detectors and a Bouncing Rubber Ball

David Marasco
Foothill College

A bouncing rubber ball under a motion sensor is a classic of introductory physics labs. It is often used to measure the acceleration due to gravity, and can also demonstrate conservation of energy. By observing that the ball rises to a lower height upon each bounce, posing the question "what is the main source of energy loss?" and requiring students to construct their own measured values for velocity from position data, a rich lab experience can be created that results in good student discussions of proper analysis of data, and implementation of models. The payoff is student understanding that seemingly small differences in definitions can lead to very different conclusions.

This experiment is a spin on the classic "measure the acceleration due to gravity" lab that uses rubber balls and motion detectors[1,2], and the follow-up experiments that look at conservation of energy[3,4]. In a previous lab, students were asked to find g using a racquetball and a motion detector. One observation they made was that as each additional bounce was at a lower height, the ball was losing mechanical energy. With that lab as prior experience, students are asked to collect position data, determine velocity information, and then make a graph that will help them answer the question "what is the dominant mode of energy loss for the bouncing ball?"

The students are told to collect data at 50 Hz and to export their position vs. time data into a spreadsheet, and rather than using the instrument-generated values for measured velocity, students are reminded that average velocity is defined as

$$V = \Delta x / \Delta t. \tag{1}$$

Most students build their velocity data by taking the position and time from one row in their spreadsheets and subtracting that from the row below. This velocity can be squared and divided by two to represent a normalized kinetic energy (mass is not included in this case as it also part of the potential energy, so a conservation of energy experiment need not include a measurement of the mass). The interesting problem at the heart of this experiment occurs when students try to add their kinetic energy data with the potential energy data they derive from their position as a function of time to create an overall ball energy.

In general, the class will divide into two groups. Many of the students (labelled Group One) will take the following approach to determining velocity, their first row of velocity data in their spreadsheet will be on the second row of their position data, as follows:

$$V_n = (x_n - x_{n-1}) / (t_n - t_{n-1}) \tag{2}$$

Each row afterwards will reference the current position and the immediately previous position. When combined with a normalized potential energy of g*h, this will generate the chart shown in figure 1, this shows a steady decay in total energy, suggesting drag as the main source of energy loss, as energy is being lost in flight as opposed to the bounces.

Most of the rest of the students (Group Two) will opt for a slightly different approach, while the first velocity will still occur on the second row, it will reference positions from the second and third rows.

$$V_n = (x_{n+1} - x_n) / (t_{n+1} - t_n) \tag{3}$$

This approach generates a graph as in figure 2, suggesting that mechanical energy is lost on the bounces. And that energy increases while the ball is in flight between bounces!

After generating their graphs, the Group One students start writing up their results, claiming that drag is the mechanism for energy loss, while the Group Two cohort are either attributing the loss to bouncing, or (hopefully) scratching their heads, wondering why the total energy is increasing from the time the ball leaves the ground to just before it strikes it again. What the instructor should do at this point is encourage students to talk to their neighboring groups, selecting pairings with opposing results. Students will claim that the other teams are either "bouncing the ball wrong" or collecting data incorrectly. The students should be encouraged to email each other their raw position vs. time data, so each team can closely examine the other's work. The students are in for a surprise, the other people's data gives the examining students' initial results!

What the students discover is that both groups have an issue with the way average velocity, and therefore kinetic energy, is being calculated[5]. The formulation $v = \Delta x/\Delta t$ best represents the velocity at the middle of the time interval, and the two different options presented are assigning that velocity to one of the endpoints. The potential energy is being calculated based on position at step n, whereas the kinetic energy is being evaluated at either step n+1/2 or n-1/2, so when the potential and kinetic energies are summed, the total energy is not a good representation of the system at a given time.

The solution is to "line up" the kinetic energy in time with the potential energy. The simplest way to do this is to take the difference in position and time from the rows immediately above and below:

$$V_n = (x_{n+1} - x_{n-1}) / (t_{n+1} - t_{n-1}) \tag{4}$$

As $t_n$ is the midpoint between $t_{n+1}$ and $t_{n-1}$, the average velocity is correctly assigned for this step. Students can determine that the main source of energy loss is the bounce, as seen in figure 3.

There are many lessons for students in this exercise. Small changes in the way velocity is calculated lead to wildly different answers to the stated question of "where does the energy get lost?" When comparing solutions, groups will argue which has the correct approach, when neither will give the solution that makes the most physical sense. Finally, many students are reticent to double the size of their time interval, as they have been told that the smaller the time interval, the closer the value of the average velocity will be to the instantaneous velocity. While this is true (and is readily apparent in the data taken for this experiment when sampled at 25 and 100 Hz), the key problem of the value at the middle of the interval needs to be properly addressed.

The author would like to thank Natasha Holmes for discussions on the further development of this lab.

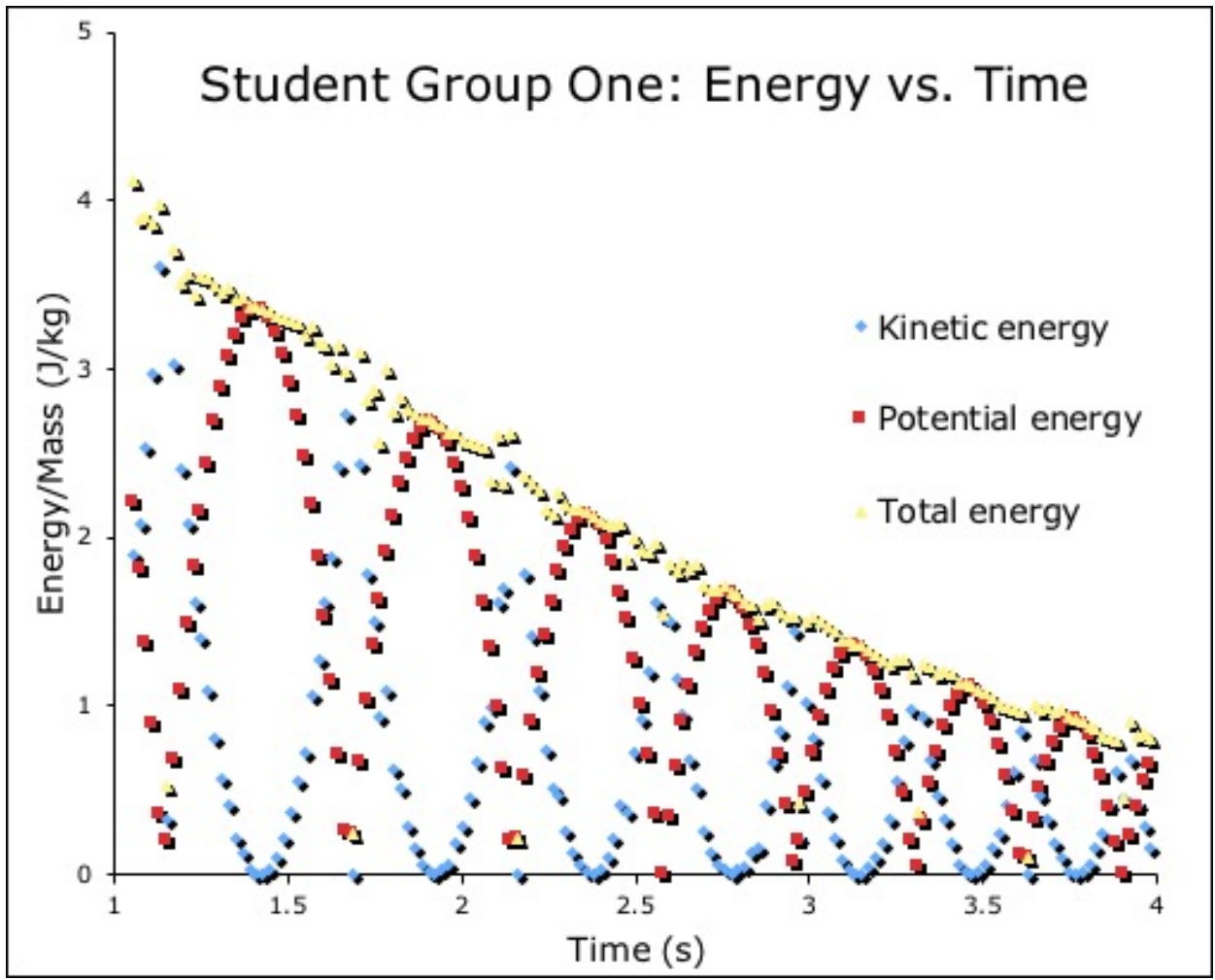

Figure 1 - The first attempt by many students will show a continual drop in total energy, pointing to a loss of energy via drag.

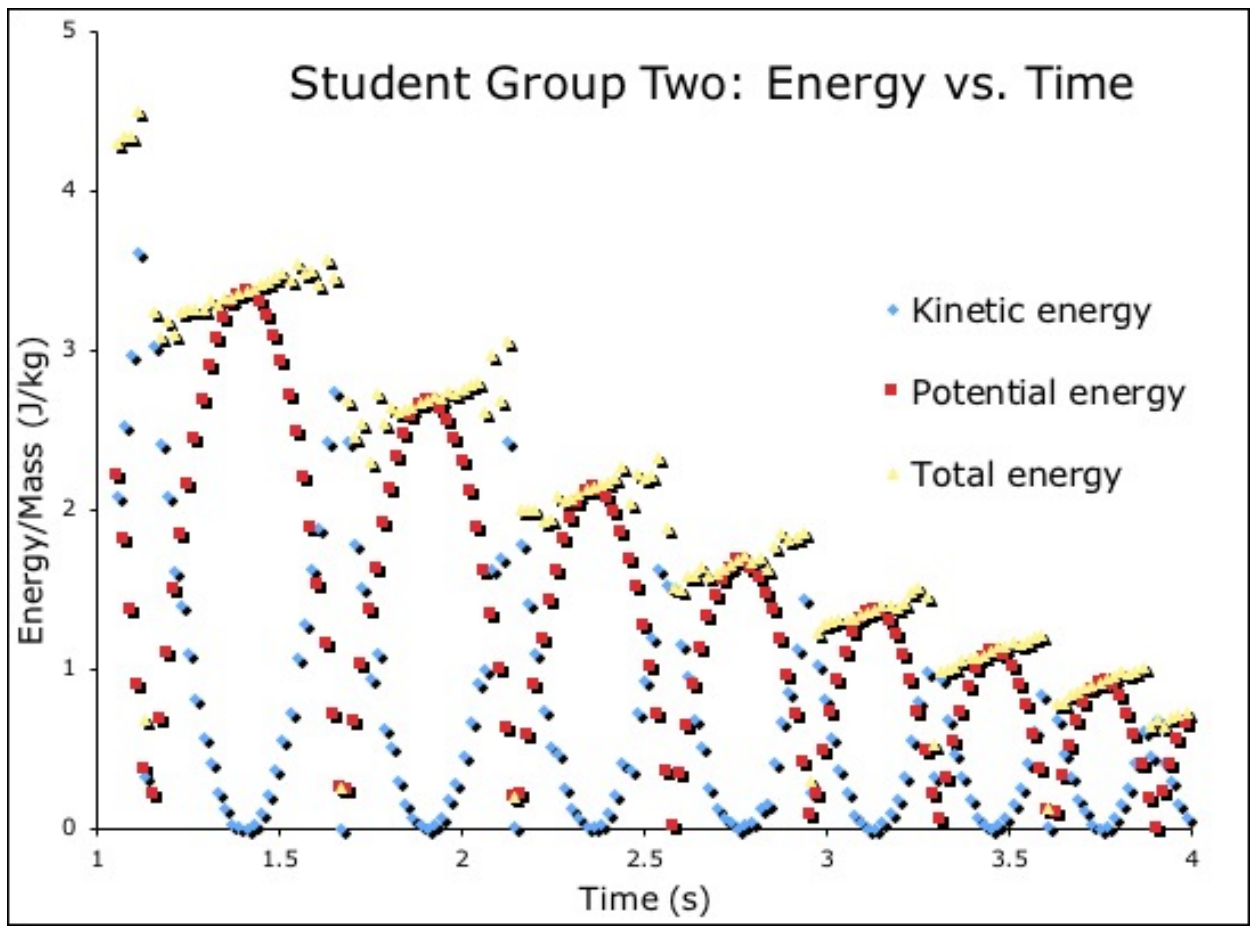

Figure 2 - Other students will model the data so that energy is lost when the ball hits the floor, but mysteriously increases in between bounces.

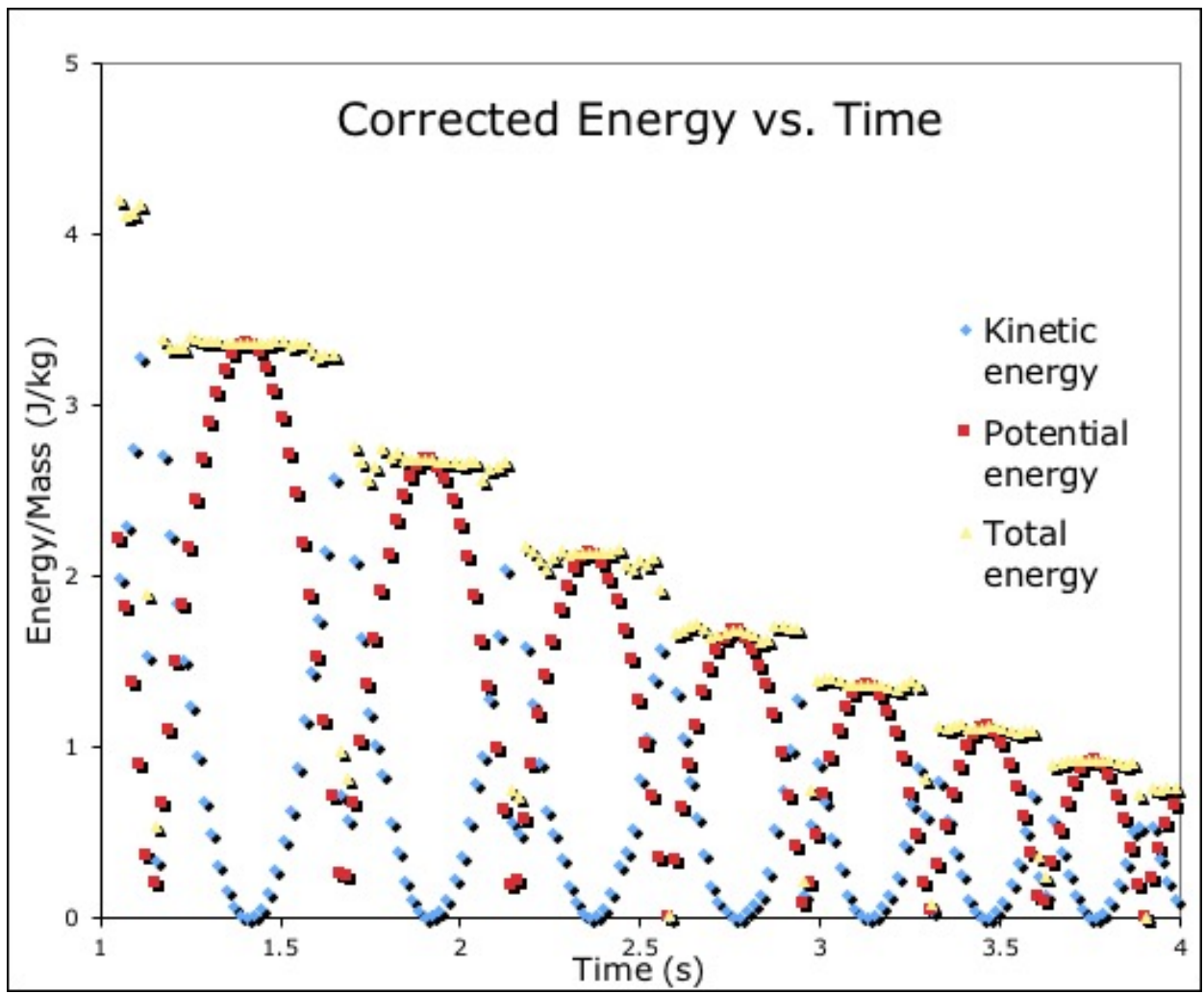

Figure 3 - Students will discover how to properly evaluate their velocities, and will find that energy is lost when the ball strikes the floor.